\documentclass[sigconf]{acmart}

\usepackage{color,soul}
\usepackage{booktabs}
\usepackage[most]{tcolorbox}
\usepackage{xcolor}
\usepackage{fontawesome5}

\usepackage{draftwatermark}

\SetWatermarkText{PREPRINT}
\SetWatermarkScale{1}
\SetWatermarkColor[gray]{0.92}

\AtBeginDocument{%
  }

\setcopyright{acmlicensed}
\copyrightyear{2026}
\acmYear{2026}
\acmDOI{XXXXXXX.XXXXXXX}

\acmConference[ACCESS-MI'26]{Context-Aware Assistive Agents for Accessible Computing}{October 05--09, 2026}{Naples, Italy}
\acmISBN{978-1-4503-XXXX-X/2026/10}

\begin{document}


\title{From Plots to Words: Model-Aware Multimodal Explanations as a Foundation for Accessible, Non-Visual Interaction}


\author{Nur Kele\c{s}o\u{g}lu}
\email{nkelesoglu@iitis.pl}
\orcid{0000-0002-0306-7281}
\affiliation{%
  \institution{Institute of Theoretical and Applied Informatics, PAS}
  \city{Gliwice}
  \country{Poland}
}

\author{Łukasz Sobczak}
\email{lsobczak@iitis.pl}
\orcid{0000-0001-9439-1812}
\affiliation{%
  \institution{Institute of Theoretical and Applied Informatics, PAS}
  \city{Gliwice}
  \country{Poland}
}

\author{Joanna Domańska}
\email{joanna@iitis.pl}
\orcid{0000-0002-1935-8358}
\affiliation{%
  \institution{Institute of Theoretical and Applied Informatics, PAS}
  \city{Gliwice}
  \country{Poland}
}

\renewcommand{\shortauthors}{Kele\c{s}o\u{g}lu et al.}

\begin{abstract}
Multimodal large language models are increasingly used in interactive systems, yet ensuring consistent, trustworthy reasoning across heterogeneous modalities remains challenging. We present a context-aware, multi-agent framework that integrates textual queries, numerical data, visual representations, and model-derived signals for explainable time-series forecasting. A distinctive feature is that it turns predominantly \emph{visual} forecasting outputs (e.g., trend plots) into structured, model-aware \emph{textual} explanations. We argue that this makes the approach a natural foundation for \emph{non-visual, accessible interaction of particular relevance to blind and visually impaired users, for whom plot-centric interfaces are largely inaccessible.} The framework supports three progressively richer pipelines (baseline, interpretable, explainable), enabling systematic comparison of unimodal, perception-driven, and model-aware responses. In an exploratory evaluation using an LLM-based judge as an early-stage proxy for human assessment, the explainable configuration improves overall explanation quality by up to 32\% over a numerical baseline, with notable gains in trustworthiness and model awareness. We position user-centered validation with target users, including screen-reader and speech-interface users, as the essential next step rather than a claim established here.

\end{abstract}




\begin{teaserfigure}
  \begin{center}
  \includegraphics[width=0.77\textwidth]{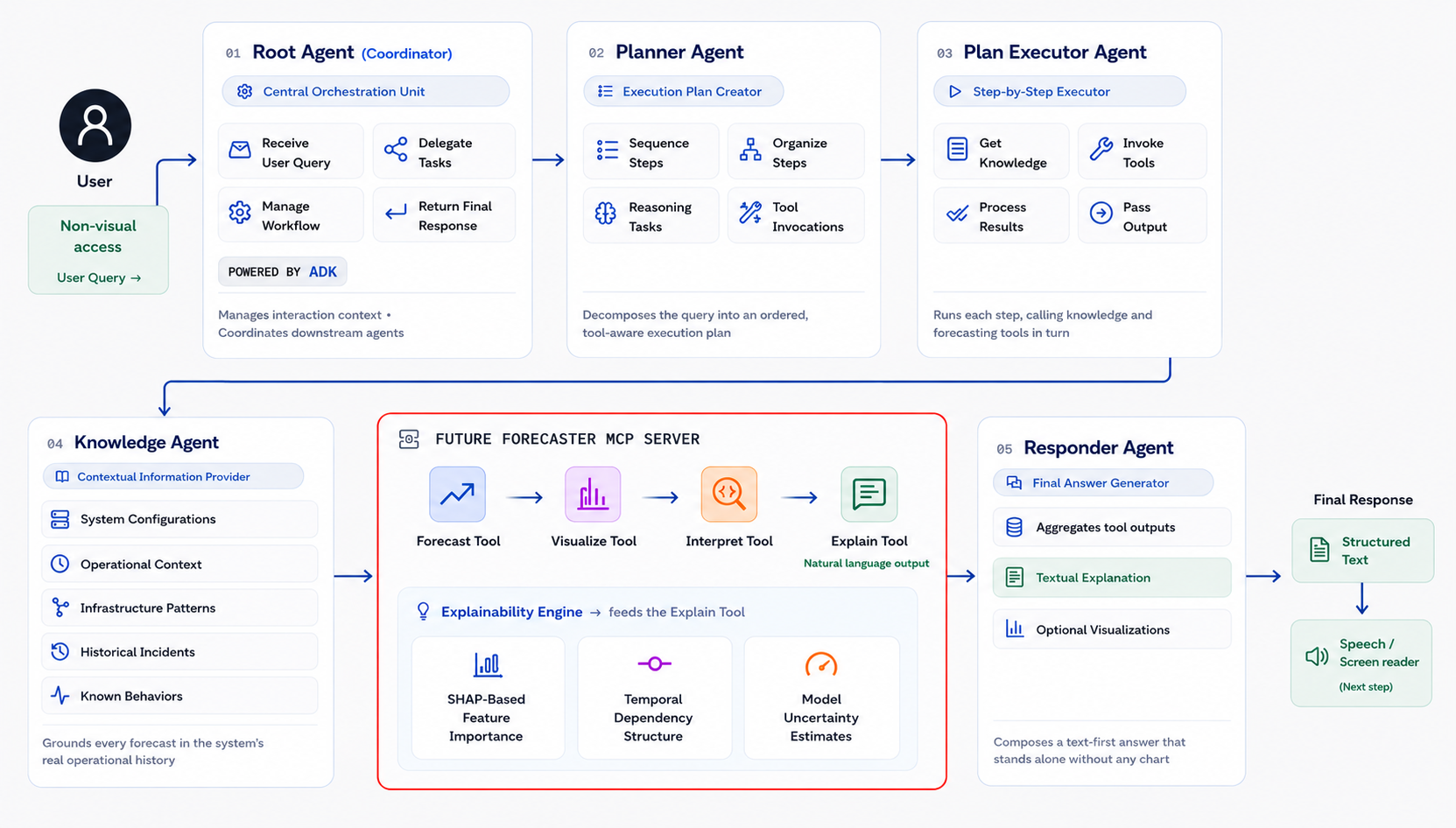}
   \end{center}
  \caption{Overview of the proposed multi-agent framework for multimodal, explainable forecasting and accessible non-visual interaction} 
  \label{fig:architecture}
\end{teaserfigure}


\maketitle

\begingroup
\small
\noindent\textbf{Preprint.}
This is the author's preprint of a paper accepted for publication in the
Proceedings of the \textit{ACCESS-MI: Context-Aware Assistive Agents for Accessible Computing}
workshop at the \textit{ACM International Conference on Multimodal Interaction (ICMI 2026)}.
The final authenticated version will appear in the ACM Digital Library.
\par
\endgroup

\vspace{0.75em}


\section{Introduction}

Recent advances in large language models (LLMs) have transformed the design of intelligent systems, enabling natural language interaction, reasoning, and decision support. Increasingly, these models are extended to multimodal settings that combine textual reasoning with structured data, visual representations, and model-derived signals such as feature attributions and uncertainty estimates. This is especially relevant to accessible computing: because LLMs can express complex, heterogeneous information as natural language, they are increasingly used as assistive agents that let users, including blind and low-vision users, interact with otherwise visual interfaces through language~\cite{kodandaram2024savant}, 
including making data graphics queryable in natural language \cite{seechart2023}. This positions multimodal LLMs (MLLM) as context-aware interaction engines for non-visual, human-centered access to data-intensive systems.

Such data-intensive systems are exactly where these needs converge, and we ground our study in one representative setting: DevOps and cloud infrastructure management. Here, large-scale time-series data must be continuously monitored to support workload forecasting, anomaly detection, and resource allocation. While modern forecasting models achieve high accuracy, their outputs are often difficult to interpret and rarely provide explanations that are both user-centered and faithful to model behavior, leaving operators with limited insight into why predictions are made and reducing trust in decision-making.

A further, often overlooked limitation is that forecasting outputs are consumed through predominantly \emph{visual} interfaces---trend plots, dashboards, and charts. For blind and low-vision users, and in eyes-free settings, these are largely inaccessible, and screen readers convey little of the meaning carried by a forecast curve or an uncertainty band~\cite{lundgard2021accessible}. Prior work makes charts accessible through natural-language interfaces that let blind users query them conversationally~\cite{murillo2020audial, seechart2023}, but targets static, general-purpose charts rather than model-aware forecasting in operational settings. Rendering forecasting behavior as structured, model-aware \emph{textual} explanations is therefore not only more interpretable but also inherently better suited to \emph{non-visual} interaction, since such outputs are modality-agnostic and can be delivered through screen readers or speech. We adopt this accessibility perspective as an explicit motivation, while treating validation with target users as future work.

This has motivated interest in explainability (XAI), which provides post-hoc justifications for \textit{\textbf{why}} a decision was produced, and interpretability (IAI), which exposes \textit{\textbf{how}} a decision is derived \cite{vishwarupe2022explainable}. In time-series settings, explanations typically come either from language-based reasoning, where LLMs describe trends, or from post-hoc analytical methods such as LIME \cite{ribeiro2016should}, SHAP \cite{lundberg2020local}, ANCHORS \cite{ribeiro2018anchors}, and saliency-based methods \cite{simonyan2013deep}. These are often disconnected: language-based reasoning lacks grounding in model behavior, analytical methods are not integrated into interactive workflows, and most approaches fail to reason jointly over numerical data, visualizations, and model-derived signals, limiting usability in practice.

In this paper, we propose a multimodal, multi-agent interaction framework for explainable time-series forecasting in DevOps systems, designed to convey model behavior through language rather than plots. Rather than treating explainability as a post-hoc step, we integrate model-aware signals, SHAP-based feature attribution, temporal dependencies, and uncertainty estimates, directly into an interactive reasoning pipeline whose textual, uncertainty-explicit outputs are inherently suited to non-visual, accessible interaction. We generate responses through three progressive pipelines: a baseline using only numerical forecasts, an interpretable setting that adds visualizations, and an explainable setting whose model-aware signals remain accessible independently of any chart. A cross-modal grounding strategy enforces consistency across numerical, visual, and model-derived signals, reducing hallucination and improving the faithfulness users rely on when they cannot inspect the data visually.
We evaluate the system on a large-scale GPU cluster dataset, showing that multimodal grounding improves explanation quality while model-aware signals further enhance trustworthiness and insightfulness. We further argue that expressing model behavior as structured, uncertainty-explicit text
makes such systems a natural foundation for accessible, non-visual interaction in Section~\ref{sec:accessibility}.

The rest of the paper is organized as follows. Section~\ref{sec:related_works} reviews related work on accessible non-visual interaction, LLM-based time-series analysis, and multimodal explainability. Section~\ref{sec:system_overview} presents the proposed system architecture. Section~\ref{sec:Experimental_setup} describes the experimental setup, including the dataset, preprocessing steps, forecasting model, and evaluation methodology. Section~\ref{sec:results} reports the experimental results. Section~\ref{sec:accessibility} discusses the framework from an accessibility and non-visual interaction perspective. Finally, Section~\ref{sec:conclusion} concludes the paper and outlines directions for future work.
\vspace{-3mm}
\section{Related Works}
\label{sec:related_works}

In this section, we first review MLLMs for accessible, non-visual interaction. Then, we turn to our application area, reviewing LLM-based time-series analysis and the ways explainability is realized in LLM-driven multimodal systems, and highlight the gap our work addresses.

A growing body of work studies multimodal LLMs as assistive agents for accessible, non-visual interaction. Recent systems augment screen readers with LLMs so that blind users can operate heterogeneous application interfaces through natural language: Savant~\cite{kodandaram2024savant}, for example, automates tedious screen-reader actions from flexible spoken commands, and related efforts combine LLMs with computer vision to add semantic context and non-linear navigation for blind and low-vision users. A particularly relevant thread targets \emph{non-visual access to data graphics}, precisely the barrier that plot-centric forecasting interfaces create. Natural-language interfaces such as AUDiaL~\cite{murillo2020audial} and SeeChart \cite{seechart2023} let blind users query charts conversationally, and Lundgard and Satyanarayan \cite{lundgard2021accessible} show that the \emph{semantic level} of a description (from low-level encodings up to high-level trends and model-relevant insights) strongly shapes its usefulness, with richer, insight-level content often preferred. Multimodal authoring tools like Umwelt~\cite{zong2024umwelt} further combine textual description, sonification, and visualization rather than merely translating visual features. Separately, HCI studies of conversational agents report that higher explainability increases user trust and acceptance \cite{joshi2024explainability}. Finally, accessibility is not language-neutral: large-scale analyses of the multilingual web show that assistive technologies frequently misrender non-Latin scripts and that native-language accessibility metadata is often missing \cite{langcrux2025}, underscoring the cultural and linguistic dimension emphasized by this workshop. These findings motivate our design: model-aware, uncertainty-explicit \emph{textual} explanations map naturally onto the semantic levels shown to matter for non-visual users, yet such explanations have not been brought to interactive forecasting in operational, cloud/DevOps settings, where outputs remain overwhelmingly visual.

We now turn to our application area. A recent line of work has explored LLMs for time-series analysis by reformulating temporal data into natural language representations \cite{kong2025position}, enabling tasks such as time-series interpretation and question answering, including trend detection, seasonal pattern analysis, similarity evaluation, and causal reasoning \cite{chen2025mtbench, kong2025time, wang2025chattime, xie2025chatts}. While such methods improve the accessibility and flexibility of time-series analysis, they rely primarily on language-based reasoning and do not provide model-aware or quantitatively grounded explanations of forecasting behavior.

Explainability in these LLM-driven multimodal systems is realized through four recurring strategies. The first two are \emph{textual} and \emph{report-oriented} explanation. Early efforts generate natural-language rationales alongside predictions: MLlm-DR \cite{zhang2025mllm} produces clinician-style rationales for depression recognition, and TimeXL \cite{jiang2025timexl} couples an LLM with a prediction--reflection--refinement loop to improve both forecasting and interpretability. Report-oriented systems instead aggregate modalities into human-readable summaries, as in MMRepAgent \cite{li2025mmrepagent}, which integrates market data, news events, and predictive models into financial reports with text, visualizations, and knowledge graphs. While readable, these explanations remain largely implicit and presentation-oriented, without exposing the internal behavior of the underlying model. The remaining two strategies pursue \emph{causal or knowledge-grounded} and \emph{intrinsic or concept-based} explanation. HEAL-LLMF \cite{haq2026heal} combines multimodal fusion with Bayesian causal inference for rationale-aligned diagnosis, and DiagLLM \cite{wang2025diagllm} integrates signal representations with expert knowledge for knowledge-grounded reasoning; both add structured reasoning yet still offer limited visibility into fine-grained model behavior such as feature-level contributions or uncertainty. In parallel, non-LLM concept-based frameworks for large multimodal models \cite{parekh2024concept} interpret predictions through human-understandable semantic concepts, improving semantic transparency but with little integration of temporal modeling or uncertainty-aware analysis. Across all four strategies, explanations are rarely grounded in quantitative, model-internal signals and are seldom designed for the interaction channel through which a user actually consumes them.

In contrast to prior work, our approach provides a comprehensive, model-aware explainability framework for multimodal time-series forecasting. Unlike LLM-based methods that rely primarily on textual or knowledge-driven reasoning, we integrate analytical and data-driven explanation components, including: SHAP-based feature attribution for fine-grained contribution analysis, temporal dependency structures to capture sequential dynamics, forecast behavior analysis across time horizons, model uncertainty estimation for reliability assessment, and multimodal visual representations combined with LLM-based reasoning. 
Equally important for expressing model behavior as structured, uncertainty-explicit text rather than as visual artifacts, our framework connects this model-aware explainability to accessible, non-visual interaction, a link that, to our knowledge, prior forecasting systems in cloud/DevOps settings have not addressed.
\vspace{-3mm}
\section{System Overview}
\label{sec:system_overview}

The proposed system is a multimodal, multi-agent framework designed to support explainable time series forecasting. It integrates LLMs with external analytical tools exposed via the Model Context Protocol (MCP), enabling both predictive capabilities and human-understandable explanations. The system operates as an interactive chatbot, where user queries are processed through a coordinated pipeline of specialized agents. Because every response is produced as structured natural-language text rather than as a rendered chart, the interaction layer is modality-agnostic by design, which makes it a natural fit for non-visual, accessible access to forecasting outputs as we discuss in Section~\ref{sec:accessibility}.

The architecture is implemented as a multi-agent system built on top of an Agent Development Kit (ADK), which provides standardized abstractions for agent communication, tool invocation, and context management. Within this framework, individual agents are powered by LLMs, with GPT-5-mini \cite{openaiOpenAIPlatform} serving as the default reasoning model across the system, offering a balance between performance and computational efficiency in multi-step interaction pipelines.

\vspace{-3mm}
\subsection{Architectural Overview}
Figure~\ref{fig:architecture} illustrates the overall architecture of the proposed system. The system operates as an interactive pipeline where user queries are processed through a hierarchy of specialized agents, enabling both structured reasoning and context-aware response generation.

The architecture follows a hierarchical multi-agent design, where responsibilities are decomposed into distinct roles. At the highest level, the system consists of six main components: \textbf{Root Agent, Planner Agent, Plan Executor Agent, Knowledge Agent, Future Forecaster MCP Server, and Responder Agent.}

These components interact sequentially but allow for iterative refinement, especially in conversational settings where the user may ask follow-up questions. The system workflow begins with a user query, which is passed to the Root Agent. The query is then transformed into an execution plan, processed step-by-step, enriched with multimodal outputs (e.g. plots, metrics), and finally translated into an explainable response.

\vspace{-3mm}
\subsection{Root Agent }

The Root Agent serves as the central orchestration unit of the system. Its primary responsibilities include: Receiving and interpreting the user query, Delegating tasks to downstream agents,  Managing the overall execution flow, and Returning the final response to the user.
The Root Agent maintains a global interaction context, which is critical for handling multi-turn conversations and ensuring coherence between successive queries. It coordinates communication between agents using ADK-provided interfaces, enabling structured message passing and tool invocation.

\vspace{-3mm}
\subsection{Planner Agent}

The Planner Agent is responsible for transforming the user query into a structured execution plan. This plan is expressed as a sequence of steps, each corresponding to specific tool invocations or reasoning tasks.

The planning process is context-aware and leverages: \textit{Knowledge of available tools (e.g., forecasting, visualization, explanation), Predefined plan templates for common DevOps scenarios, The semantic interpretation of the user query.}

In addition to dynamic plan generation, the Planner Agent supports predefined execution templates for recurring and well understood use cases. These templates encode domain knowledge and best practices, enabling faster and more reliable plan construction in scenarios such as anomaly investigation, resource utilization analysis, or forecasting-based diagnostics. When a user query matches a known pattern, the agent can instantiate and adapt a predefined plan rather than generating it entirely from scratch, improving both efficiency and consistency of the system’s behavior.

For example, a query regarding abnormal GPU energy consumption may result in a plan consisting of: (1) Loading historical metrics, (2) Performing forecasting, (3) Visualizing trends, and (4) Generating an explanation. The generated plan is exposed to the user before execution, increasing transparency and enabling user trust, which aligns with the overall goal of providing interpretable and context-aware system behavior.

\vspace{-3mm}
\subsection{Plan Executor Agent}

The Plan Executor Agent is responsible for executing the plan generated by the Planner Agent. It operates in a step-by-step manner, where each step may involve: Getting additional data from the Knowledge Agent, Calling external tools, Processing intermediate results, and Passing outputs to subsequent steps.

Through ADK integration, the executor can dynamically invoke MCP tools and manage their inputs and outputs in a structured way. A key characteristic of this component is its ability to integrate multimodal outputs. For instance, it can combine numerical forecasts with visual representations and pass both to downstream components.

\vspace{-3mm}
\subsection{Knowledge Agent}

The Knowledge Agent is responsible for providing structured domain knowledge about the monitored system, which can be used to enrich both interpretation and explanation processes. Its primary role is to supply contextual information that is not directly observable in the raw time-series data but is essential for accurate and meaningful reasoning. This includes, for example: \textit{System configuration details} (e.g., cluster size, resource limits), \textit{Operational context} (e.g., deployment schedules, maintenance windows), \textit{Known system behaviors and constraints}, and \textit{Historical incidents and typical patterns observed in the infrastructure.}

From an architectural perspective, the Knowledge Agent integrates seamlessly with the multi-agent framework and can be invoked by the Plan Executor when required by the execution plan. The retrieved knowledge is then propagated to downstream components, particularly the explanation module, where it is combined with multimodal inputs such as historical data, forecasts, and visualizations.

\vspace{-2mm}
\subsection{Future Forecaster MCP Server}
\label{sec:mcp_server}
The Future Forecaster MCP Server encapsulates domain-specific analytical capabilities and exposes them as modular tools that can be dynamically invoked by the Plan Executor Agent. In contrast to conventional monolithic pipelines, the MCP-based design enables flexible composition of forecasting, visualization, interpretation, and explainability functionalities. The server provides four primary tools:

\begin{itemize}
    \item \textbf{Forecast Tool} – generates future predictions based on historical time-series data using a machine learning model (Random Forest with autoregressive lag features). In addition to forecast outputs, this tool exports intermediate artifacts (model, features, inputs) required for downstream explainability.

    \item \textbf{Visualize Tool} – produces both static and dynamic visual representations of historical and forecasted data. This includes comparison plots and animated simulations, enabling temporal understanding of forecast evolution.

    \item \textbf{Interpret Tool} – generates \textit{data-driven multimodal interpretations} by jointly analyzing historical data, forecast outputs, and visual representations. This tool leverages a multimodal LLM to provide descriptive explanations of trends, seasonality, and anomalies, but does not incorporate model-level reasoning.

    \item \textbf{Explain Tool} – generates \textit{model-aware, context-grounded explanations} by integrating: Historical and forecast data, Visual representations, SHAP-based feature importance, Temporal dependency structure, and Model uncertainty estimates.
  
    This tool enforces structured reasoning constraints to ensure that explanations explicitly reference model behavior rather than relying solely on observable trends. 
    To enable access to SHAP-based feature importance, temporal dependencies, and model uncertainty, the Explain Tool integrates outputs from the Explainability Engine, which computes model-specific explainability signals.
    
\end{itemize}

\paragraph{Progressive Forecasting Result Pipelines}

A key design contribution of this work is the introduction of a three-level response generation framework that progressively increases multimodal context and model awareness:

\begin{enumerate}
    \item \textbf{Baseline Forecasting (BF)}  
      Generated using a unimodal language model (\textit{GPT-5 mini}) that receives only forecasted numerical values as input. This configuration represents a unimodal pipeline, where the language model receives only numerical forecast outputs. 

    \item \textbf{Interpretable Forecasting (IF)}  
    This pipeline introduces multimodal grounding by incorporating visual representations (e.g., forecast plots) alongside textual inputs. The model can leverage temporal patterns, trends, and anomalies observable in the data, leading to improved interpretability, produced by a multimodal LLM (\textit{GPT-5.2}) that incorporates both textual and visual inputs. However, the reasoning remains perception-driven and does not explicitly reflect the internal behavior of the forecasting model.

    \item \textbf{Explainable Forecasting (EF)}  
    This pipeline extends multimodal reasoning with explicit model-awareness. In addition to numerical and visual inputs, the system integrates model-derived signals such as SHAP-based feature attributions, temporal importance, and uncertainty estimates. This enables explanations that are grounded not only in observed data but also in the decision-making process of the predictive model.  Generated by the proposed framework using a multimodal LLM (\textit{GPT-5.2}) augmented with model-aware signals.
\end{enumerate}

This progressive design enables a systematic comparison across three levels of reasoning: \textit{forecast outputs} (BF), \textit{perception-level explanations} (IF), and \textit{model-aware explanations} (EF). It highlights how increasing multimodal grounding and explainability enhances the quality, reliability, and trustworthiness of generated responses.

A central challenge in multimodal LLM systems is ensuring consistency across heterogeneous modalities, including numerical data, visual representations, and textual explanations. In the proposed framework, both Interpretable Forecasting (IF) and Explainable Forecasting (EF) pipelines leverage multimodal inputs, combining textual data with visual representations (e.g., forecast plots) to support richer reasoning.

The IF pipeline achieves \textit{perception-level grounding}, where explanations are derived from observable patterns in historical and forecasted data, supported by statistical summaries and visual trends. While this enables coherent and context-aware interpretations, the reasoning remains limited to patterns that can be directly inferred from the data.

In contrast, the EF pipeline extends this by introducing \textit{model-aware grounding}, where explanations must additionally align with internal model signals, including SHAP-based feature attributions, temporal importance, and uncertainty estimates. This enforces a stronger form of cross-modal consistency, where generated explanations must simultaneously reflect: Quantitative forecast outputs, Visual trends and temporal patterns, Model-internal signals (feature importance and uncertainty).

\vspace{-2mm}
\subsection{Responder Agent}

The Responder Agent is responsible for synthesizing the final answer presented to the user. It aggregates: Outputs from all executed plan steps, generated visualizations, and explanation text from the MCP server. Based on these inputs, it constructs a coherent and contextually relevant response aligned with the user’s original query. The response may include both textual explanations and references to visual artifacts, ensuring that the final output remains both informative and accessible.

\vspace{-2mm}
\section{Experimental Setup}
\label{sec:Experimental_setup}
This section describes the experimental setup used to evaluate the proposed framework, including the dataset, preprocessing steps, forecasting model, and evaluation methodology.

\vspace{-2mm}
\subsection{Dataset}

The GPU Cluster Spot Resource Dataset \cite{ovi_gpu_dataset_analysis}, provides a large-scale, production-grade trace of AI workloads executed on a heterogeneous GPU cluster supporting both high-priority and spot workloads. The infrastructure consists of 4,278 GPU nodes and 11,702 GPU cards across multiple GPU types. The dataset tracks 466,867 job submissions from 119 organizations over approximately 113 days, capturing job-level metadata such as requested GPUs, CPUs, worker counts, job duration, and submission timestamps. This makes it highly suitable for large-scale workload characterization and predictive system behavior modeling.

From a system behavior modeling perspective, this dataset is particularly valuable for developing forecasting models for GPU demand, understanding concurrency patterns, and evaluating capacity planning strategies. The mixture of heterogeneous GPU types and multi-organization competition provides a realistic benchmark for predictive DevOps resource management. Its rich, multi-dimensional metadata also makes it a suitable testbed for studying how such complex operational signals can be conveyed to users through language rather than visual dashboards, in line with the non-visual interaction focus of this work.
\vspace{-2mm}
\subsection{Data Preprocessing}
To enable time-series forecasting of GPU demand, the raw dataset was filtered to extract the target metric, namely \textit{gpus\_active\_requested}, which represents the number of actively requested GPUs over time.

A supervised learning formulation was adopted by converting the time series into a lag-based feature representation. Specifically, a sliding window approach was used to generate 24 lag features, where each instance is constructed from the previous 24 time steps to predict the current GPU demand. This effectively captures short-term temporal dependencies and workload dynamics in the cluster.

The dataset was then split into training and test sets using a hold-out strategy, where the last H observations (forecast horizon) were reserved for evaluation, and the remaining data was used for training.

\vspace{-2mm}
\subsection{Forecasting Model Training and Explanability Tools}

To model GPU demand, a Random Forest Regressor was employed due to its robustness in capturing non-linear relationships and temporal dependencies in tabular time-series representations. The model was trained using the lag-based feature matrix, where the previous 24 time steps serve as input to predict the current GPU demand. The training process uses 100 decision trees with a fixed random seed to ensure reproducibility. After training, the model is serialized and stored for subsequent inference and explainability analysis.

For the explainability component, SHAP (TreeExplainer) is used with a background sample of 100 instances from the training data to approximate feature contributions. Feature importance is aggregated across time and further grouped into: Recent (lag 1--3), Mid-term (lag 4--12), and Long-term (lag 13+). Model uncertainty is estimated by computing the standard deviation of predictions across ensemble trees for each forecast step.

\vspace{-2mm}
\subsection{Recursive Forecasting Formulation}
For forecasting, a recursive multi-step strategy was adopted. Starting from the most recent 24 observed values, the model predicts the next time step, and this prediction is iteratively fed back as input to generate future forecasts. This process continues for the entire forecasting horizon, enabling flexible multi-step prediction without requiring a separate model per horizon. During inference, all intermediate input vectors used for prediction are recorded and stored, facilitating downstream analysis such as explainability and feature attribution.

\vspace{-2mm}
\subsection{Agents and Large Language Model}

The system employs large language models as the core reasoning engine within the multi-agent architecture. Two configurations are used:

\begin{itemize}
    \item \textbf{GPT-5-mini}: Used by the Planner, Knowledge Agent, Plan Executor and Responder agents for efficient reasoning and orchestration.
    
    \item \textbf{GPT-5.2}: Used in the \textit{Interpret} and \textit{Explain} tools for multimodal reasoning, due to its stronger capabilities in handling joint textual and visual inputs.
\end{itemize}

The LLM operates in a multimodal setting, where inputs may include: \textit{ Structured numerical data} (historical and forecasted time series), \textit{Visual representations} (forecast plots), and \textit{Model-derived signals} (SHAP values and uncertainty estimates).

For the \textit{Explain} tool, a low-temperature configuration ($T=0.2$) is used to ensure deterministic and stable outputs, particularly important for model-aware explanations. The maximum output length is set to 1500 tokens to allow structured, multi-section responses.

\paragraph{Multimodal Reasoning Setup}

To ensure cross-modal consistency, the LLM is prompted with explicitly structured inputs that combine:  Data-level information (time series and statistical summaries), Visual evidence (encoded forecast graphs), and Model-level explanations (SHAP and uncertainty). This design enforces grounded reasoning, where generated explanations must align with all available modalities, reducing hallucination and improving interpretability.

\vspace{-2mm}
\subsection{Environment}

The system is deployed as a modular, containerized architecture consisting of a frontend, a multi-agent reasoning layer, and a backend analytical layer. The frontend enables natural language interaction and visualization of forecasts and explanations.

Reasoning is handled by a multi-agent system using the OpenAI API, where GPT-5-mini is used for orchestration and GPT-5.2 for multimodal interpretation and explanation. Analytical capabilities are exposed via an MCP server implemented with FastMCP, providing access to forecasting and explainability tools.

The forecasting module is based on a Random Forest model (scikit-learn), with feature attribution computed using SHAP, and data processing performed with pandas, NumPy, and Matplotlib. The containerized setup ensures reproducibility and a clear separation between reasoning and computation.

\vspace{-2mm}
\subsection{User Queries}

To evaluate the proposed system in realistic interaction scenarios, we constructed a set of 30 representative user queries reflecting typical DevOps tasks related to GPU cluster monitoring, time-series analysis, and operational decision-making.

The queries cover a broad spectrum of use cases, including detecting overload risks and capacity violations, analyzing temporal patterns (e.g., seasonality, spikes, variability) and identifying anomalies and assessing system stability. They also contain forecasting long-term trends for capacity planning, reasoning under uncertainty and recommending mitigation or optimization strategies. Each query is formulated as a multi-step analytical request requiring the system to combine forecasting, interpretation of time-series behavior, and decision support, ensuring evaluation in a realistic, context-rich interaction setting.

\vspace{-2mm}
\subsection{Evaluation}

We evaluate the proposed system by comparing three pipelines: 
(i) a baseline forecasting pipeline (BF), 
(ii) an interpretability-enhanced pipeline (IF), and  (iii) the proposed  xplainability-enhanced pipeline (EF). For evaluation, we construct a dataset of 30 representative user queries related to GPU cluster monitoring, time-series forecasting, and system behavior analysis. Each query is processed by all three pipelines under identical conditions, producing three corresponding responses per query.

\vspace{-2mm}
\subsubsection{LLM-based Evaluation Protocol}

Due to the subjective and multi-dimensional nature of explanation quality, we adopt an LLM-based evaluation protocol using the latest GPT-5.4 language model as an automatic judge. This approach has been increasingly used as a proxy for human evaluation in recent work on generative models.

The evaluator model is provided with the original user query, the system response, and a description of the system and its operational constraints (e.g., GPU cluster capacity and scheduling behavior).

The model is instructed to critically assess each response according to a set of predefined criteria reflecting both user-centric and model-centric aspects of explanation quality. To reduce variance, all evaluations are performed with deterministic decoding (temperature = 0), and identical prompts are used across all system variants.

\begin{table*}[t]
\centering
\begin{tabular}{lcccccccccc}
\toprule
\textbf{System} & \textbf{Clarity} & \textbf{Help.} & \textbf{Insight} & \textbf{Trust} & \textbf{Model} & \textbf{Cons.} & \textbf{Hall.} $\downarrow$ & \textbf{Act.} & \textbf{Uncert.} & \textbf{Overall} \\
\midrule
Baseline Forecasting & 3.93 & 3.57 & 2.97 & 2.40 & 1.63 & \textbf{3.93} & 1.00 & 1.87 & 1.80 & 2.69 \\
Interpretable Forecasting & \textbf{4.00} & \textbf{4.50} & 3.77 & 2.97 & 3.03 & 3.73 & 0.97 & \textbf{2.00} & 1.80 & 3.29 \\
Explainable Forecasting & \textbf{4.00} & 4.33 & \textbf{3.87} & \textbf{3.10} & \textbf{4.37} & 3.77 & \textbf{0.90} & 1.97 & \textbf{2.00} & \textbf{3.56} \\
\bottomrule
\end{tabular}
\caption{Quantitative comparison of system variants across evaluation metrics. Higher is better for all metrics except hallucination. Best results are highlighted in bold.}
\label{tab:metrics_results}
\end{table*}

\subsubsection{Evaluation Metrics}

We assess each response using the following metrics:

\paragraph{Clarity (0--5)}
Measures how understandable and well-structured the explanation is. A high score indicates that the response is easy to follow and does not require additional interpretation effort.
\vspace{-2mm}
\paragraph{Helpfulness (0--5)}
Evaluates whether the response directly addresses the user’s query and provides relevant information for the task at hand.
\vspace{-2mm}
\paragraph{Insightfulness (0--5)}
Captures the extent to which the explanation provides non-obvious, meaningful insights beyond surface-level description, such as identifying underlying patterns or causes.
\vspace{-2mm} 
\paragraph{Trustworthiness (0--5)} Assesses whether the explanation increases confidence in the prediction, for example by providing justification, reasoning, or supporting evidence.
\vspace{-2mm}
\paragraph{Model Awareness (0--5)}
Measures whether the explanation reflects the underlying model behavior, including references to features, temporal dependencies, or explainability mechanisms.
\vspace{-2mm}
\paragraph{Consistency (0--5)}
Evaluates the internal logical coherence of the response, ensuring that statements are not contradictory and follow a consistent reasoning flow.
\vspace{-2mm}
\paragraph{Hallucination (0--1)}
Indicates whether the response contains unsupported or unrealistic claims. A value of 1 denotes the presence of hallucinated content.
\vspace{-2mm}
\paragraph{Actionability (0--2)}
Measures whether the response provides concrete, actionable recommendations that can be applied in a DevOps context.
\vspace{-2mm}
\paragraph{Uncertainty Awareness (0--2)}
Assesses whether the response acknowledges prediction uncertainty and communicates limitations of the forecast.
\vspace{-2mm}
\subsubsection{Aggregation}

For each system variant, scores are averaged across all queries. Additionally, we compute a weighted overall score that emphasizes insightfulness, trustworthiness, and model awareness, as these dimensions are central to explainability.

\vspace{-2mm}
\section{Results}
\label{sec:results}

This section presents a comparative evaluation of the proposed multimodal explainable forecasting framework against two baseline configurations. The goal is to assess how increasing levels of multimodal context and model-awareness affect the quality, usefulness, and reliability of generated responses. All resources required to reproduce our results, including source code, user queries, generated responses, and evaluation scripts, are publicly available at \textit{\url{https://zenodo.org/records/19710740}.} 

We evaluate three pipelines (BF, IF, EF) defined in Section~\ref{sec:mcp_server}. For each user query, we generate and analyze three types of responses.
We evaluate all configurations using a set of user-centric and model-centric metrics, including clarity, helpfulness, insightfulness, trustworthiness, model awareness, consistency, hallucination rate, actionability, and uncertainty awareness.

\begin{figure}[t]
\centering
\includegraphics[width=0.9\linewidth]{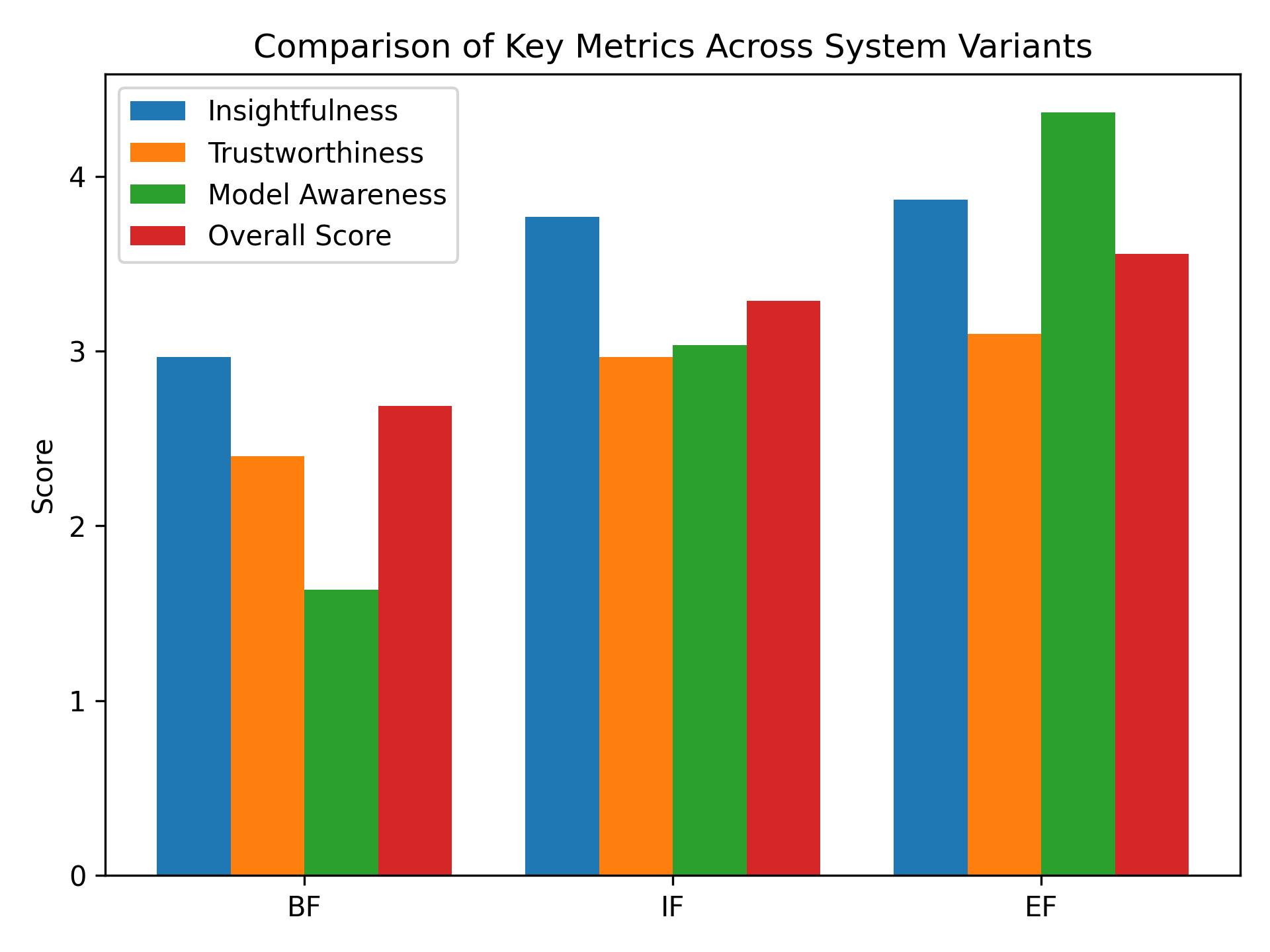}
\caption{Comparison of key explainability-related metrics across system variants.}
\label{fig:bar_chart}
\end{figure}

Table~\ref{tab:metrics_results} summarizes the aggregated results across all queries. IF improves the overall score from 2.69 to 3.29, mainly due to gains in helpfulness, insightfulness, and model awareness. EF achieves the highest score (3.56), with further improvements in model awareness, trustworthiness, and insightfulness, while reducing hallucination and improving uncertainty awareness, with comparable clarity and consistency. To further illustrate these differences, Figure~\ref{fig:bar_chart} presents a comparison of key explainability-related metrics, including insightfulness, trustworthiness, model awareness, and overall score. The figure clearly shows a monotonic improvement from BF to IF and EF, with the largest gains observed in model awareness and insightfulness.

\begin{figure}[t]
\centering
\includegraphics[width=0.9\linewidth]{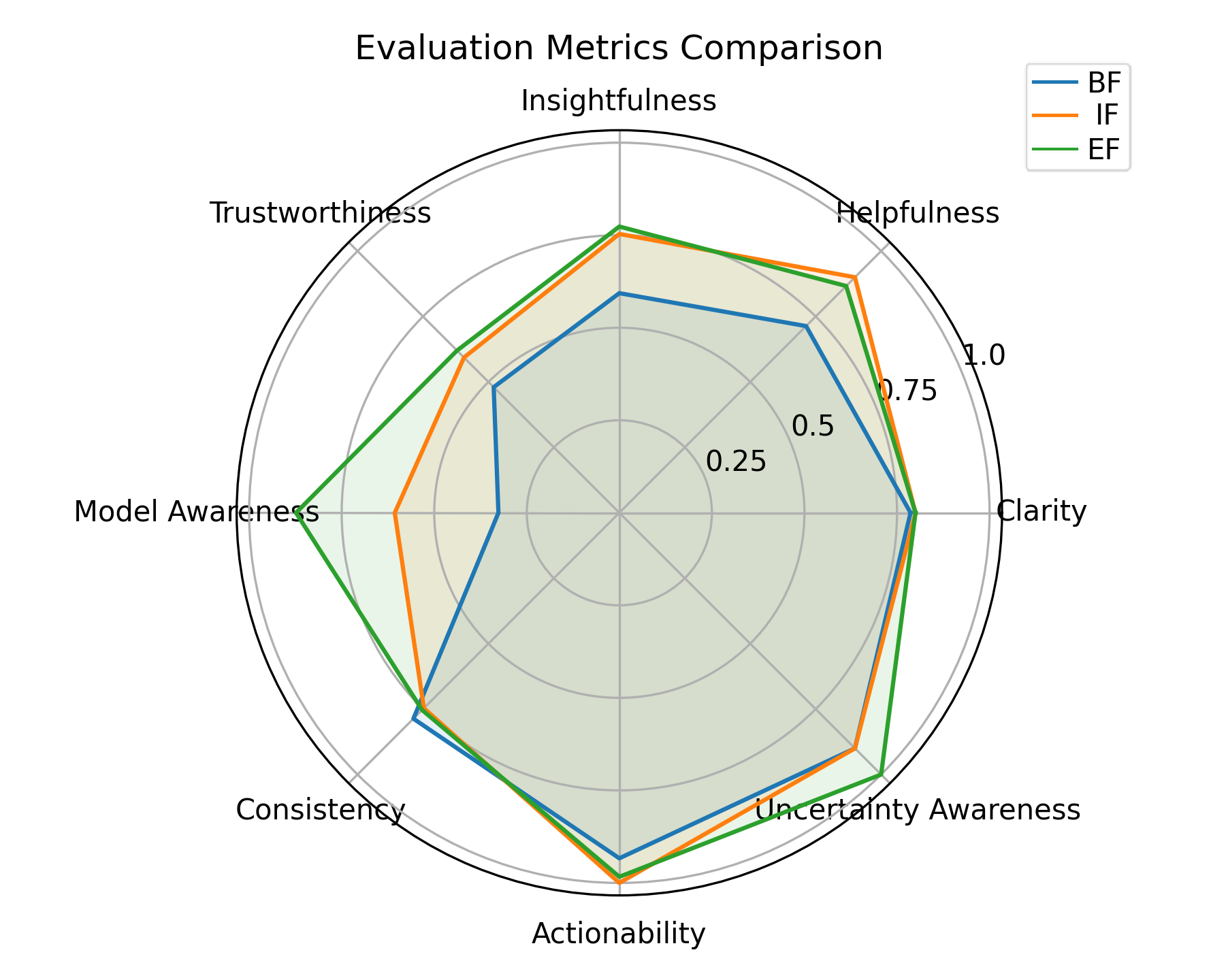}
\caption{Normalized radar plot illustrating the performance profile of each system across evaluation metrics. All metrics are scaled to the [0,1] range for comparability.}
\label{fig:radar_plot}
\end{figure}

Additionally, Figure~\ref{fig:radar_plot} provides a holistic view of system performance across all evaluation dimensions. The radar plot highlights the balanced performance of the EF configuration, which consistently outperforms or matches the baselines across most metrics, while maintaining high clarity and consistency. Notably, the radar plot reveals that improvements in explainability do not come at the cost of usability, as clarity and consistency remain stable across configurations. The results demonstrate a clear and consistent improvement as additional multimodal and explainability components are introduced. In particular, the explainable forecasting configuration achieves the highest overall score, indicating that incorporating model-aware signals significantly enhances the quality of generated explanations. 

We observe that multimodal grounding (IF) substantially improves helpfulness and insightfulness compared to the baseline, confirming the importance of visual and contextual information in interpreting time-series forecasts. However, the largest gains are observed when explicit explainability signals are introduced.

Notably, EF achieves the highest scores in insightfulness, trustworthiness, and model awareness, indicating that explanations grounded in model behavior provide more meaningful and reliable insights for users. Additionally, EF improves uncertainty awareness and reduces hallucinations, suggesting that model-aware reasoning leads to more faithful and robust responses.

To quantify the observed improvements, we compute the relative gains in overall performance between system variants. The interpretable forecasting configuration improves the overall score by approximately \textbf{22\%} compared to the baseline, highlighting the benefits of multimodal grounding. Further incorporating explicit explainability mechanisms yields an additional improvement of approximately \textbf{8\%} over IF. 

Overall, the proposed explainable forecasting framework achieves a total improvement of approximately \textbf{32\%} compared to the baseline, demonstrating the cumulative impact of multimodal context and model-aware explanations. These results indicate that multimodal grounding contributes the majority of the performance gains, while explicit explainability provides additional improvements, particularly in model awareness and trustworthiness.
\vspace{-2mm}
\section{Accessibility Perspective and Non-Visual Interaction}
\label{sec:accessibility}
A property of our framework that matters for accessible computing is that it converts predominantly visual forecasting outputs into structured, model-aware textual explanations; we offer this as a position and design analysis, not an empirical accessibility result. Plot-centric interaction, reading trend lines and inferring uncertainty from shaded bands, is inaccessible to blind and low-vision users and in eyes-free settings, since screen readers cannot recover a chart's semantics. This distinguishes our pipelines: the Interpretable (IF) pipeline is perception-driven and assumes visual access, whereas the Explainable (EF) pipeline grounds explanations in model-derived signals (SHAP attributions, temporal importance, uncertainty) available independently of any image. EF can thus convey why a forecast behaves as it does and how confident the model is entirely in language, making it a natural basis for non-visual interaction, where explicit textual uncertainty becomes a prerequisite for trustworthy decisions. Being modality-agnostic, these outputs could also drive a speech-based, eyes-free loop without changing the core pipeline, though such a loop is not implemented or evaluated here. We are deliberate about scope: the system was not designed with or evaluated by target users, and our LLM-based judge is only an early-stage proxy. We therefore claim not accessibility but a promising foundation for it, and regard participatory evaluation with blind and low-vision users as the essential next step.

\vspace{-2mm}
\section{Conclusion}
\label{sec:conclusion}
In this work, we presented a context-aware, multimodal LLM-based framework for accessible, non-visual interaction with time-series forecasting, delivering model behavior entirely through language rather than the plots and dashboards that exclude blind and low-vision users. Rather than treating explainability as a post-hoc add-on, the system integrates forecasting, visualization, and model-aware reasoning within a unified multi-agent architecture, producing structured, uncertainty explicit explanations that are grounded in model behavior and therefore usable without any visual channel,a natural foundation for non-visual, accessible interaction.

Our findings show that multimodal grounding improves clarity, helpfulness, and insightfulness, while explicit model-aware signals such as feature attributions, temporal dependencies, and uncertainty estimates, further enhance trustworthiness and model awareness. The progressive comparison across baseline, interpretable, and explainable pipelines highlights that multimodal perception improves understanding, but true explainability requires alignment with the model's internal decision process. Conceptually, by separating interpretation from explanation and enforcing cross-modal grounding, our approach addresses the disconnect between fluent language generation and faithful reasoning, positioning LLMs as context-aware interaction engines rather than mere text generators.

Several challenges remain. The framework's effectiveness depends on the quality and alignment of multimodal inputs, and while grounding reduces hallucinations, it does not eliminate them, indicating a need for stronger verification. A further limitation is that our evaluation relies on an LLM-based judge as an early-stage proxy that cannot substitute for human assessment, so user-centered properties such as real trustworthiness and usability are not yet established empirically. Future work will explore adaptive explanation strategies for different user expertise levels, tighter integration of causal reasoning, formal validation of explanation faithfulness, and extensions to additional domains and real-time settings. In addition, we plan participatory evaluations with blind and low-vision users, and intend to realize and evaluate a speech-based, non-visual interaction loop around the existing model-aware explanation layer, the most direct route from the foundation laid here to a genuinely accessible assistive system.

Overall, this work demonstrates that explainable forecasting in real-world systems is inherently a multimodal interaction problem, and that combining LLM-based reasoning with model-aware signals and cross-modal grounding is a key step toward trustworthy, human-centered, and accessible AI systems.

\vspace{-2mm}
\section*{Safe and Responsible Innovation Statement}
This work explores multimodal LLM-based decision support for DevOps systems using system-level, non-personal telemetry data, and supports human decision-making rather than full automation. Our evaluation involves no human participants and relies on an LLM-based judge, so we make no empirical claims about accessibility or user experience—a limitation we state explicitly (Sections~\ref{sec:accessibility} and~\ref{sec:conclusion}). Framing this work in an accessibility context carries a specific responsibility: assistive explanations must not create a false sense of reliability for users who cannot independently inspect the data. We mitigate over-reliance through explicit communication of uncertainty and model limitations, and any deployment for blind or low-vision users would require participatory design and evaluation before accessibility can be claimed.

\vspace{-2mm}
\section*{Acknowledgments}

This work was partly supported by the National Centre for Research and Development (NCBR) and co-funded by the European Union under the European Funds for Modern Economy (FENG) Programme (SMART Konsorcja), project no.\ FENG.01.01-IP.01-A0GV/24-00, ``Advanced maintenance and diagnostic tool for IT start-ups''.

\bibliographystyle{ACM-Reference-Format}
\bibliography{References}

\end{document}